\documentclass[useAMS,usenatbib]{mn2e}
\usepackage[pdftex]{graphicx}
\usepackage{xcolor}
\usepackage{subfigure}
\usepackage{amsmath}
\usepackage{amsfonts}
\usepackage{amssymb}

\newenvironment{indentpar}[1]%
 {\begin{list}{}%
         {\setlength{\leftmargin}{#1}}%
         \item[]%
 }
 {\end{list}}

\newcommand{\kms}{km s$^{-1}$}
\def\q#1{`#1'}

\newcommand{\deaf}{`d', `e' and `f'}

\newcommand{\cd}{\q{d}}
\newcommand{\ce}{\q{e}}
\newcommand{\cf}{\q{f}}
\newcommand{\ap}{$\sim$ }

\title[The Conversion of Gas to Stars in YMCs]{Tracing the Conversion of Gas into Stars in Young Massive Cluster Progenitors}
\author[D. L. Walker et al.]{D.~L.~Walker$^{1}$\thanks{E-mail: \texttt{D.L.Walker@2009.ljmu.ac.uk}}, 
S.~N.~Longmore$^{1}$,
N.~Bastian$^{1}$,
J.~M.~D.~Kruijssen$^{2}$,
\and J.~M.~Rathborne$^{3}$ ,
J.~M.~Jackson$^{4}$, 
J.~B.~Foster$^{5}$
and Y.~Contreras$^{3}$\vspace{0.5cm}  \\
$^{1}$Astrophysics Research Institute, Liverpool John Moores University, IC2, 146 Brownlow Hill, Liverpool, L3 5RF, United Kingdom\\
$^{2}$Max-Planck Institut fur Astrophysik, Karl-Schwarzschild-Strasse 1, 85748, Garching, Germany \\
$^{3}$CSIRO Astronomy and Space Science, Epping, Sydney, Australia\\
$^{4}$Institute for Astrophysical Research, Boston University, Boston, MA 02215, USA\\
$^{5}$Yale Center for Astronomy and Astrophysics, New Haven, CT 06520, USA 
\\
}

\begin{document}

\date{}

\pagerange{\pageref{firstpage}--\pageref{lastpage}} \pubyear{2015}

\maketitle

\label{firstpage}
\begin{abstract}

Whilst young massive clusters (YMCs; $M$ $\gtrsim$ 10$^{4}$ M$_{\odot}$, age $\lesssim$ 100 Myr) have been identified in significant numbers, their progenitor gas clouds have eluded detection. Recently, four extreme molecular clouds residing within 200 pc of the Galactic centre have been identified as having the properties thought necessary to form YMCs. Here we utilise far-IR continuum data from the Herschel Infrared Galactic Plane Survey (HiGAL) and millimetre spectral line data from the Millimetre Astronomy Legacy Team 90 GHz Survey (MALT90) to determine their global physical and kinematic structure. We derive their masses, dust temperatures and radii and use virial analysis to conclude that they are all likely gravitationally bound -- confirming that they are likely YMC progenitors. We then compare the density profiles of these clouds to those of the gas and stellar components of the Sagittarius B2 Main and North proto-clusters and the stellar distribution of the Arches YMC. We find that even in these clouds -- {\it{the most massive and dense quiescent clouds in the Galaxy}} -- the gas is not compact enough to form an Arches-like ($M$ = 2x10$^{4}$ M$_{\odot}$, R$_{eff}$ = 0.4 pc) stellar distribution. Further dynamical processes would be required to condense the resultant population, indicating that the mass becomes more centrally concentrated as the (proto)-cluster evolves. These results suggest that YMC formation may proceed hierarchically rather than through monolithic collapse.
\end{abstract}


\section{Introduction}\
\begin{figure*}
\begin{center}
\includegraphics[scale=0.65, angle=-90]{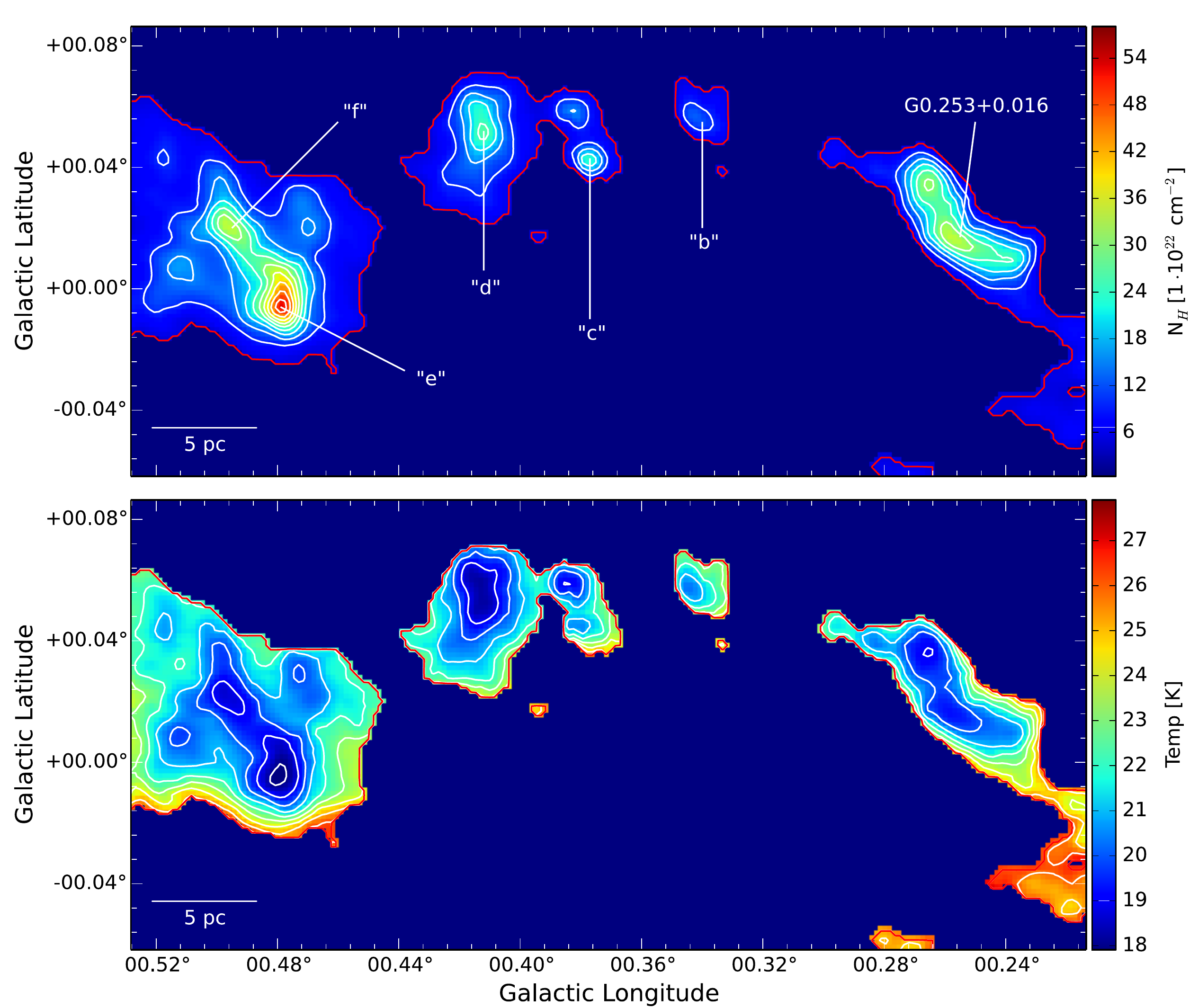}
 \caption{{\it Top}: Masked column density map of the \q{dust-ridge}, derived from HiGAL far-IR continuum observations \citep{Cara_higal}. A column density threshold of 5 x 10$^{22}$ cm$^{-2}$ (red contours) is implemented and contours (white) are at 10, 15 ... 55 x 10$^{22}$ cm$^{-2}$. {\it Bottom}: Masked temperature map of the \q{dust-ridge}, derived from HiGAL far-IR continuum observations \citep[for the techniques used to derive these maps, see][]{Cara_higal}. Contours (white) are at 17, 18 ... 27 K (systematic uncertainties in SED fitting are \ap 3--4 K). To display the correlation between dust column density and temperature, the same pixels as in the {\it{top}} image are masked here. Clearly seen in all of the clumps is an anti-correlation between column density and temperature of the dust -- consistent with them being centrally condensed with cold interiors \citep{Brick}.}
\end{center}
\end{figure*}

Stellar clusters can contribute substantially to the stellar population of a galaxy \citep[e.g.][]{Lada,embedded_survival,silva_cfe, Diederik_cfe}. Classically, they have been categorised into two distinct populations; globular clusters (GCs) and open clusters (OCs). More recently however, it has become apparent that {\it{young massive clusters}} \citep[YMCs; defined by][to be gravitationally bound systems with $M$ $\gtrsim$ 10$^{4}$ M$_{\odot}$ and ages $\lesssim$ 100 Myr]{ymc_port}, whose stellar masses and densities can reach and even exceed those of GCs, are still forming at the present day \citep[e.g.][]{Holt}. This discovery has reinvigorated the discussion of cluster formation mechanisms and spawned the idea that perhaps all clusters form in a similar way, with OCs representing the low-mass counterparts and GCs representing the early-Universe analogues of extreme YMCs \citep[][]{elm_gcs,Diederik_submitted}. Understanding and characterising the formation and evolution of YMCs may therefore be fundamental in revealing how clusters form across the full mass-range. 

YMCs with ages greater than a few Myrs have been identified in significant numbers \citep[see][and references therein]{ymc_port}, and particularly in galaxies with high star-formation rates \citep[e.g.][]{Whitmore}. Several have also been identified within our own Galaxy, such as the Arches, Quintuplet, NGC 3603, Westerlund 1 and red super-giant (RSG) clusters \citep{Figer_arches, NGC3603, Clark_west,Ben_RSG}. However, if we are to understand how such clusters form, we must identify and characterise their progenitor gas clouds. These progenitor clouds are expected to contain $\gtrsim$ 10$^{5}$ M$_{\odot}$ within only a few parsecs. Indeed, \citet{Ginsburg_clouds} and \citet{atlasgal_hii} find several such clouds in the Galactic disk that satisfy these conditions. However, these clouds are {\it{all}} actively forming stars and will therefore have lost much of their initial structure as a result of stellar feedback (note that \citet{atlasgal_hii} explicitly searched for star-forming clouds). Whilst these clouds are of course important in understanding the YMC formation process, if we are to probe the initial conditions then we require clouds that have yet to initiate widespread star formation. Thus, in addition to the aforementioned criteria, we would then also expect such progenitor clouds to be cold (T \ap 10 - 20 K) and devoid of widespread, high-mass star formation. Given that none of the identified progenitor clouds in the Galactic disk are quiescent, very little is known about the initial conditions from which these massive clusters form. Identifying and characterising the gas within such clouds is therefore vital if we are to understand the process of YMC formation. They also present a unique insight into the earliest stages of star, planet and cluster formation in extremely dense environments, where feedback mechanisms and dynamical encounters are likely to affect the subsequent cluster evolution significantly.

\citet{Brick} recently identified a likely YMC precursor in the extreme infrared dark cloud (IRDC) G0.253$+$0.016 \citep[see e.g.][]{Lis_brick1,Lis_brick2,Brick, Brick_jens, Brick_jill, Brick_KJ,Jill_pdf_2014}, and determined that it is cold (T$_{dust}$ $\sim$ 20 - 30 K), high-mass (M $\sim$ 1.3 x 10$^{5}$ M$_{\odot}$), compact (r $\sim$ 2.8 pc) and other than a single region of weak water-maser emission, exhibits little-to-no star forming activity -- precisely the conditions one would expect of a dense cloud that could form a high-mass cluster. \citet{Bricklets} later identified three further potential YMC precursors in clouds known as \q{d}, \q{e} and \q{f}. \citet{Immer} show that these clouds are all high-mass (\ap 10$^{5}$ M$_{\odot}$), compact (\ap pc-scale) and other than a region of methanol maser emission towards cloud \q{e}, are all quiescent. Along with G0.253$+$0.016, they are situated in the Central Molecular Zone \citep[CMZ; inner \ap 200 pc of the Galaxy,][]{CMZ} and belong to the so-called \q{dust-ridge} \citep{Bricklets_lis, Immer} towards the Galactic centre, which itself appears to belong to a coherent circumnuclear gas ring. \citet{Molinari_ring} model this ring as a twisted, $\infty$-shaped ellipse that orbits the central super-massive black hole, Sgr A*, with a semi-major orbital axis of \ap 100 pc. \citet{Diederik_orbit} confirm the coherence of the gas stream using dynamical modelling. The fact that four of the most extreme clouds known to exist in the Galaxy all lie at the same distance and reside within the same stream of gas is certainly very intriguing, but it also presents an opportunity to study a sample of potential YMC progenitor clouds under the same environmental conditions at similar sensitivity and resolution.

In this paper we extend the analysis of \citet{Brick} (L12, hereafter) of G0.253$+$0.016 to include clouds \deaf. We utilise continuum data from the Herschel infrared Galactic Plane Survey \citep[HiGAL,][]{higal} to measure their global physical properties such as mass, radius and temperature. These HiGAL data cover a wavelength range of 70 -- 500 $\mu$m and provide \ap 5.5" or 0.22 pc resolution at the distance of these clouds, which is taken to be \ap 8.4 kpc \citep{distance}. Spectral line data from the Millimetre Astronomy Legacy Team 90 GHz Survey \citep[MALT90,][]{malt90_foster_11,malt90_foster_13,malt90} are used to measure the global kinematic properties of these clouds. The MALT90 survey provides spectral line data for 16 lines in the 90 GHz band. The data cubes for these transitions consist of 4096 channels with 0.11 \kms \ velocity resolution. Using the results from these data, we determine whether these clouds -- the most massive and dense quiescent clouds known in our Galaxy -- are capable of forming YMCs. Having shown that they are indeed candidate YMC precursors, we then compare these clouds to more evolved (proto)-YMCs to speculate as to how gas is converted to stars on large scales in the early stages of YMC formation.

\section{Results}

\subsection{Dust Column Densities \& Temperatures}\

Figure 1 ({\it{upper panel}}) shows the HiGAL column density map of the \q{dust-ridge} at the Galactic centre. The bottom panel displays the HiGAL temperature map of the same region. \citep[See][for the techniques used to derive these maps]{Cara_higal}. A column density threshold has been applied such that all pixels below 5 x 10$^{22}$ cm$^{-2}$ are masked. This threshold is chosen as it highlights the spatial extent of the dense clumps well, whilst effectively masking the more diffuse emission across the region. G0.253$+$0.016, \deaf \ have peak column densities ranging from 2.6 -- 5.3 x 10$^{23}$ cm$^{-2}$ and central temperatures ranging 17 -- 19 K. Overall, the maps clearly display an anti-correlation between the column density and temperature of the dust in every cloud -- consistent with them being centrally condensed with cold interiors. We note that the central temperature of cloud \q{c} is slightly higher than might be expected given its central column density -- at 22 K it is \ap 4 K warmer in its core than the other clouds, despite all being at similar column densities. This is likely due to the fact that this cloud is likely forming high-mass stars that heat the cloud's interior, as evidenced by water and methanol maser emission \citep{c_watermaser, c_maser}. It is interesting to note the discrepancy between the gas and dust temperatures at the Galactic centre. \citet{Ao_gas_temp} find that the gas temperature towards the Galactic centre ranges from 50 K to in excess of 100 K. This is significantly higher than the low dust temperatures of \ap 20 K, suggesting that the gas is being heated by some non-photon driven mechanism such as cosmic ray heating or turbulent energy dissipation. With respect to our sample of clouds, we note that \citet{Ao_gas_temp} only measured the temperature of G0.253$+$0.016. We assume that their results also hold for the other three clouds.

\subsection{Dust Mass}\

Using these column density maps, we estimate the dust masses of the clouds by assuming a mean molecular weight of 2.8m$_{H}$, multiplying each pixel by its physical area (assuming a distance of 8.4 kpc) and summing over the cloud area. For G0.253$+$0.016 and cloud \q{d}, this is straightforward given the column density threshold of 5 x 10$^{22}$ cm$^{-2}$. However, clouds \q{e} and \q{f} are not entirely distinct, are embedded in a higher density region that lies above this threshold, and have two nearby distinct clumps at different velocities. Any kinematically distinct emission (identified using MALT90 data; see \S {\it{2.4}}) is therefore masked to determine the masses of clouds \q{e} \ and \q{f}. We determine the masses of G0.253$+$0.016,  \deaf \ to be 11.9 x 10$^{4}$ M$_{\odot}$, 7.6 x 10$^{4}$ M$_{\odot}$, 11.2 x 10$^{4}$ M$_{\odot}$ and 7.3 x 10$^{4}$ M$_{\odot}$, respectively (see Table 1). We note that these mass estimates are necessarily uncertain due to the ambiguity involved in defining a distinct cloud area, particularly in a contiguous region such as this.

Figure 2 shows how the estimated masses decrease as we choose higher column density thresholds. Increasing the threshold from 5 - 10 x 10$^{22}$ cm$^{-2}$ decreases our mass estimates by approximately 20\% for G0.253$+$0.016, 40\% for cloud \cd, 10\% for cloud \ce \ and 15\% for cloud \cf. Given the systematic uncertainties in estimating dense gas mass towards the Galactic centre of a factor of \ap 2 \citep{snl_sf}, any uncertainty in these mass estimates is dominated by systematics rather than the threshold used. These estimates agree well with those found by \citet{Immer} and L12. Note that the slight (\textless 10\%) discrepancy between L12's and our mass estimate for G0.253$+$0.016 is simply due to a slightly higher column density threshold.

\subsection{Radius}\

Defining characteristic cloud radii is further complicated by their non-spherical geometry. Rather than simply prescribing representative circular radii, we instead take a geometric mean of the minor and major axes of each cloud, whose boundaries are defined by the aforementioned column density threshold, to derive effective radii. We derive effective radii of 2.9 pc, 3.2 pc, 2.4 pc and 2.0 pc, respectively, for G0.253$+$0.016, \deaf. 

To compare how compact these clouds are, we determine characteristic radii within which 5 x 10$^{4}$ M$_{\odot}$ is enclosed. This mass is chosen as we know that {\it{all}} four clouds exceed this mass and it is large enough such that a 10$^{4}$ M$_{\odot}$ cluster could form from it, given a star formation efficiency as low as 20\%. Figure 3 displays this mass-radius relation for the four clouds and black dashed lines indicate their characteristic radii -- 1.0 pc, 1.2 pc, 0.8 pc and 1.0 pc, respectively for G0.253$+$0.016, \deaf. This highlights the extreme nature of these clouds in that they harbour enough mass to form a YMC within a radius of \ap 1 pc and yet they do not show any signs of high-mass star formation other than that inferred by the detection of weak maser emission. 

\begin{figure}
\begin{center}
\includegraphics[scale=0.58,angle=0]{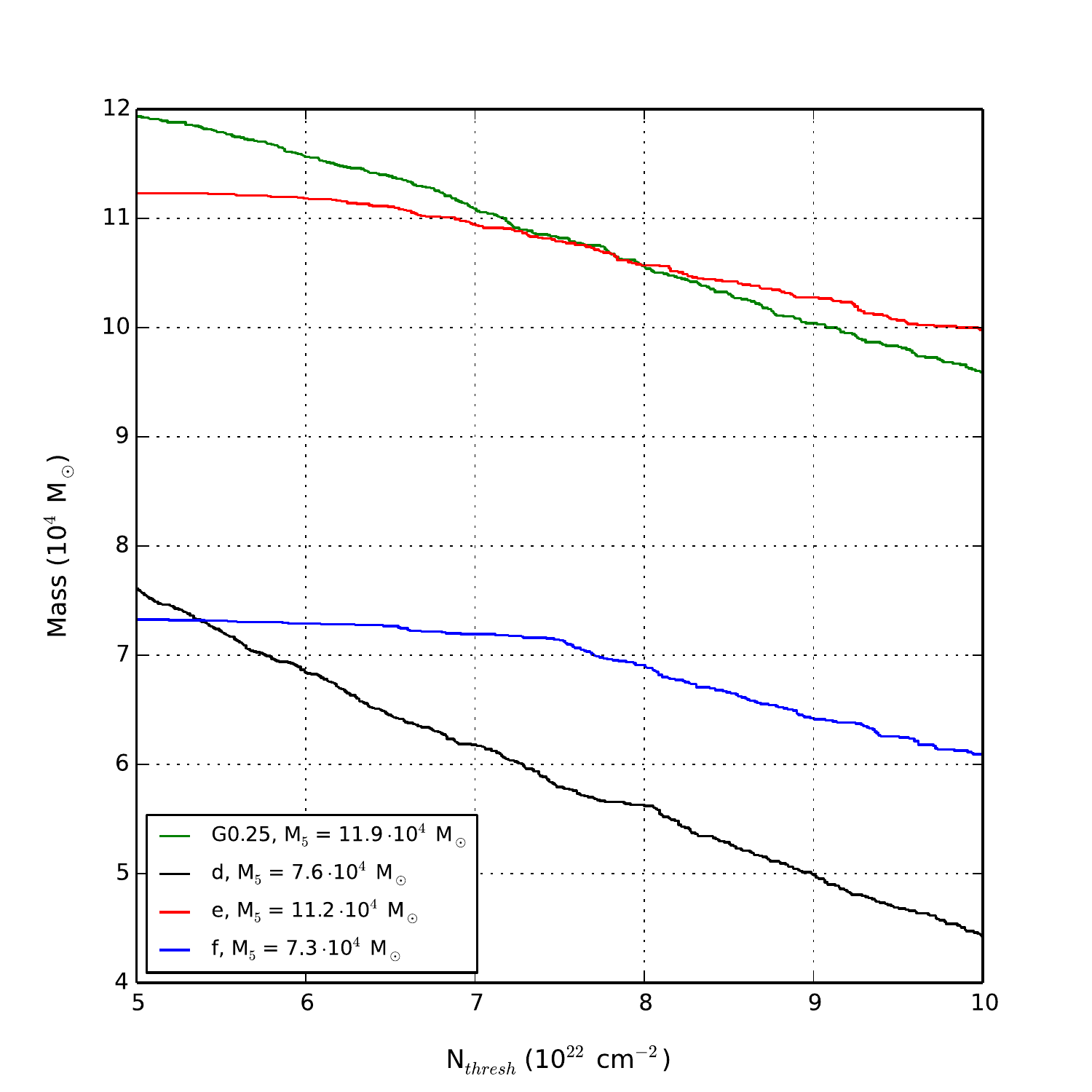} \\
 \caption{Mass determination as a function of column density threshold for clouds G0.253$+$0.016 ({\it{green}}), \cd \ ({\it{black}}), \ce \ ({\it{red}}) and \cf \ ({\it{blue}}). We implement a lower threshold of 5 x 10$^{22}$ cm$^{-2}$ and incrementally increase this to 10 x 10$^{22}$ cm$^{-2}$. Clouds \ce, \cf \ and G0.253$+$0.016 are not overly sensitive to the column density threshold, with \ap 10\% and \ap 15\% changes over this range, respectively. The mass of cloud \cd \ is however less well constrained, with a change in mass  of \ap 40\% over the range. The legend displays the mass determined for each cloud at the lowest threshold.}
\end{center}
\end{figure}

\begin{figure}
\begin{center}
\includegraphics[scale=0.58,angle=0]{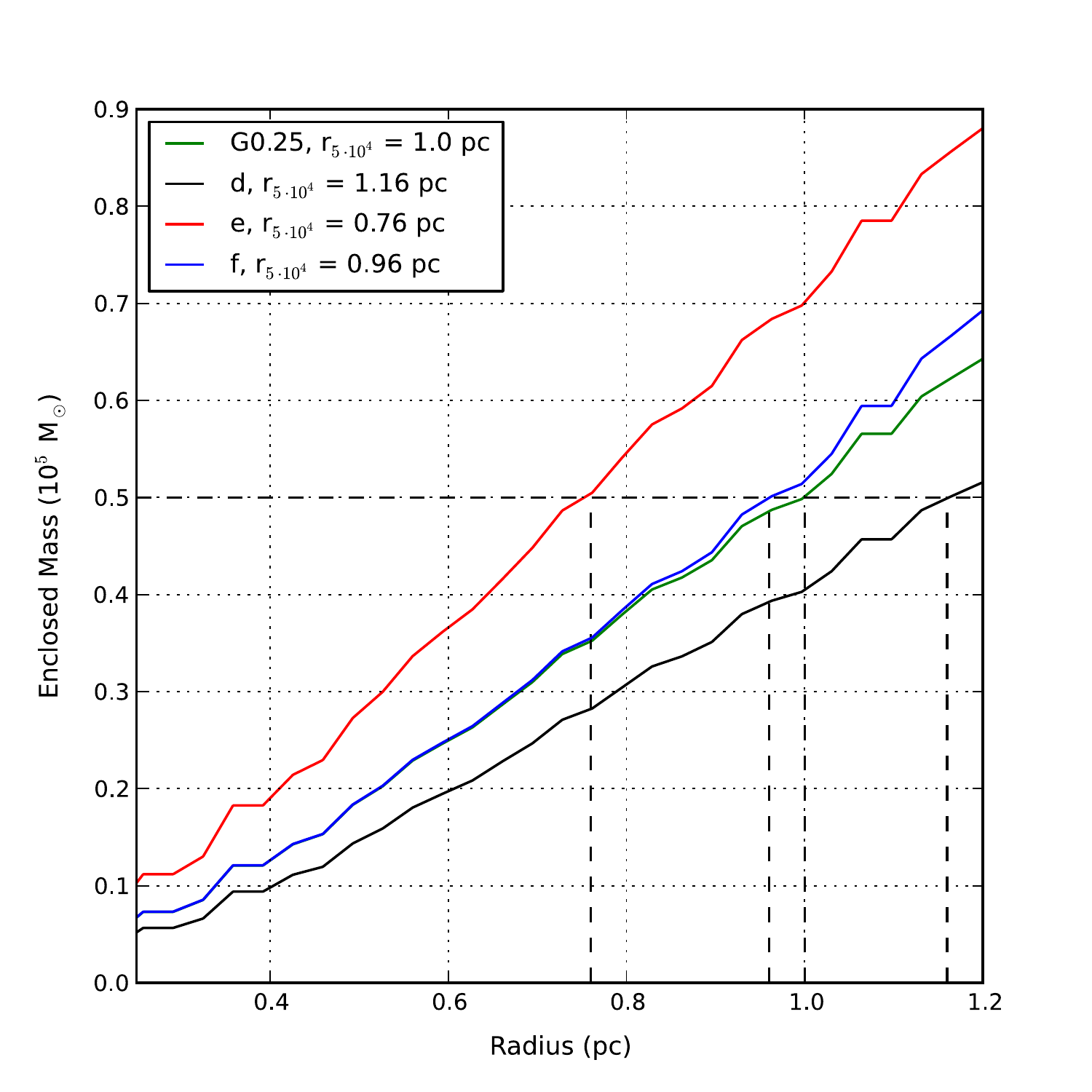} \\
 \caption{Enclosed mass as a function of radius for G0.253$+$0.016 (green), \cd \ ({\it{black}}), \ce \ ({\it{red}}) and \cf \ ({\it{blue}}). Intersections of black dashed lines indicate the radii within which 5 x 10$^{4}$ M$_{\odot}$ are enclosed (see legend).}
\end{center}
\end{figure}

\subsection{Gas Velocity Dispersion}\

We use molecular line emission to determine the global kinematics of these clouds and assess their dynamical state. This analysis is complicated due to their location in the CMZ, where the kinematic and chemical structure is known to be complex -- with large velocity dispersions, shock-enhanced chemistry and confusion due to unrelated line-of-sight emission within the Galactic disk. It is therefore important to ensure that we use sensible molecular tracers to isolate the emission from each cloud.

The MALT90 survey provides data cubes for 15 different molecular tracers, probing a range of critical densities and excitation energies, thus making it ideal for our purposes. Across clouds \deaf, emission from 11 of the 15 tracers is well detected. We adopt the same categorisation as \citet{Brick_jill}, separating these detected molecular transitions as: {\it{optically thick tracers}} -- HCN (1--0), HNC (1--0), HCO$^{+}$ (1--0) and N$_{2}$H$^{+}$ (1--0); {\it{optically thin tracers}} -- HN$^{13}$C (1--0), H$^{13}$CO$^{+}$ (1--0) and $^{13}$CS (2--1) and {\it{\q{hot core}/shock tracers}} -- HNCO 4(0,4)--3(0,3), SiO (1--0), HC$_{3}$N (10--9) and CH$_{3}$CN 5(0)--4(0). As might be expected of clouds in the turbulent environment of the CMZ, we find that their kinematics are complicated. The data cubes show complex line profiles, significant velocity gradients, multiple velocity components, large line-widths and intense shock-associated emission lines. We defer detailed analysis of these features to a subsequent paper. (See \citet{Brick_jill} for an in-depth discussion regarding the molecular line emission from G0.253$+$0.016). Here we simply wish to isolate the emission from each cloud to obtain estimates of the line-widths. To do this, we make the reasonable and common assumption that the {\it{optically thin}} transitions trace the underlying kinematics of the clouds most reliably, whereas optically thick lines only trace the surface kinematics of the clouds. Of these, HN$^{13}$C is the brightest line detected in all three clouds and is therefore used for line-width determination in all cases. Figure 4 shows Hanning-smoothed HN$^{13}$C spectra that have been averaged over the spatial extent of the clouds (defined by the 5 x 10$^{22}$ cm$^{-2}$ column density threshold, see Figure 1), where each profile is fitted using the CASA software package \citep{casa}. Cloud \cd \ ({\it{left panel}}) displays two velocity components, one at \ap 0 \kms \ and another at \ap 20 \kms. We attribute only the 20 \kms \ component to the gas in cloud \cd, as the morphology of the molecular line emission corresponds to the dust emission well. The 0 \kms \ component arises from an unrelated cloud along the line of sight (see Figure 5a). The fit to the associated component yields a peak velocity and line-width (FWHM) of V$_{LSR}$ = 20.75 $\pm$ 0.69 km s$^{-1}$ and $\Delta$V = 16.3 $\pm$ 1.5 km s$^{-1}$. Cloud \ce \ ({\it{middle panel}}) shows a clear singular component with V$_{LSR}$ = 31.66 $\pm$ 0.22 km s$^{-1}$ and $\Delta$V = 15.51 $\pm$ 0.52 km s$^{-1}$. Cloud \cf \ ({\it{right panel}}) also shows two velocity components, one at \ap 27 \kms \ and another at \ap 40 \kms. Though both components show similar line-widths and peak intensities, we conclude that only the 40 \kms \ component is associated with cloud \cf, as again the molecular line emission matches the dust emission well (see Figure 5b). This component has a peak velocity and line-width of V$_{LSR}$ = 40.1 $\pm$ 2.6 km s$^{-1}$ and $\Delta$V = 16.5 $\pm$ 3.2 km s$^{-1}$. Coupling these results with L12's results for G0.253$+$0.016 -- V$_{LSR}$ = 36.1 $\pm$ 0.4 km s$^{-1}$,  $\Delta$V = 15.1 $\pm$ 1.0 km s$^{-1}$ -- we see that all four clouds, whilst having slightly different peak velocities, have very similar line-widths of \ap 16 km s$^{-1}$.

\begin{figure*}
\begin{center}
\includegraphics[scale=0.5,angle=-90]{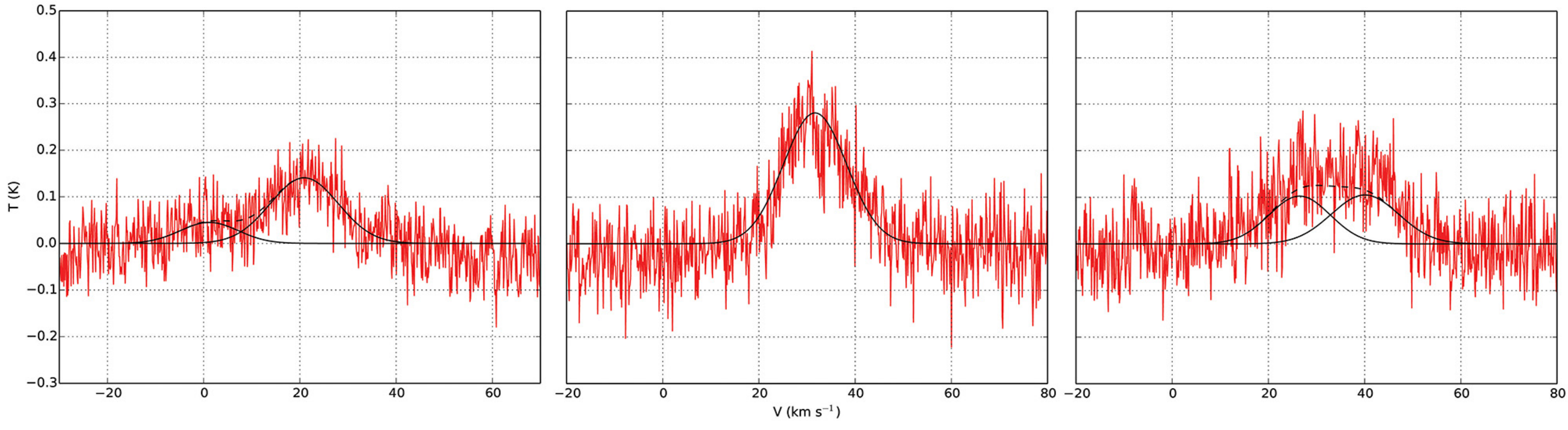} \\
 \caption{Hanning-smoothed, spatially-averaged HN$^{13}$C line profiles for clouds \q{d} \ [{\it{left}}], \q{e} \ [{\it{centre}}] \ and \q{f} [{\it{right}}]. Each is fitted using the multi-component Gaussian-fitting in the CASA software package. In the profile for cloud \cd, we attribute the component at \ap 0 \kms \ to emission from an unrelated cloud along the line of sight and in cloud \cf \ we attribute the \ap 27 \kms \ component to emission from nearby clouds (see Figure 5 for integrated intensity images). Omitting these unassociated components, we obtain the following peak velocities and line-widths: cloud {\bf d}:  V$_{LSR}$ = 20.75 $\pm$ 0.69 km s$^{-1}$, $\Delta$V = 16.3 $\pm$ 1.5 km s$^{-1}$; cloud {\bf e}:  V$_{LSR}$ = 31.66 $\pm$ 0.22 km s$^{-1}$, $\Delta$V = 15.51 $\pm$ 0.52 km s$^{-1}$; cloud {\bf f}: V$_{LSR}$ = 40.1 $\pm$ 2.6 km s$^{-1}$, $\Delta$V = 16.5 $\pm$ 3.2 km s$^{-1}$.}
\end{center}
\end{figure*}

\subsection{Virial mass -- Gravitationally bound?}\

\begin{figure}
\hfill
\subfigure[]{\includegraphics[width=4cm,angle=-90]{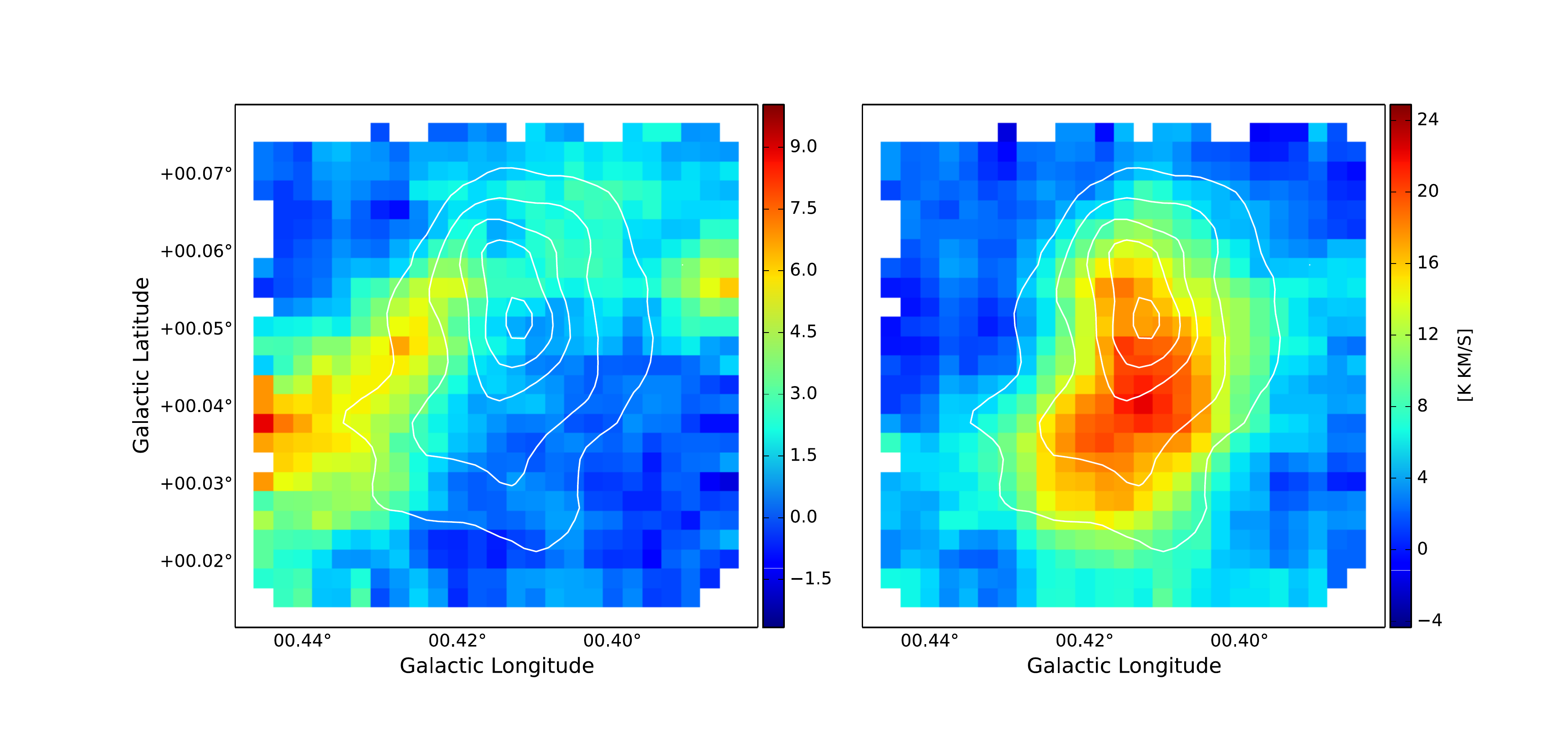}}
\hfill
\subfigure[]{\includegraphics[width=4cm,angle=-90]{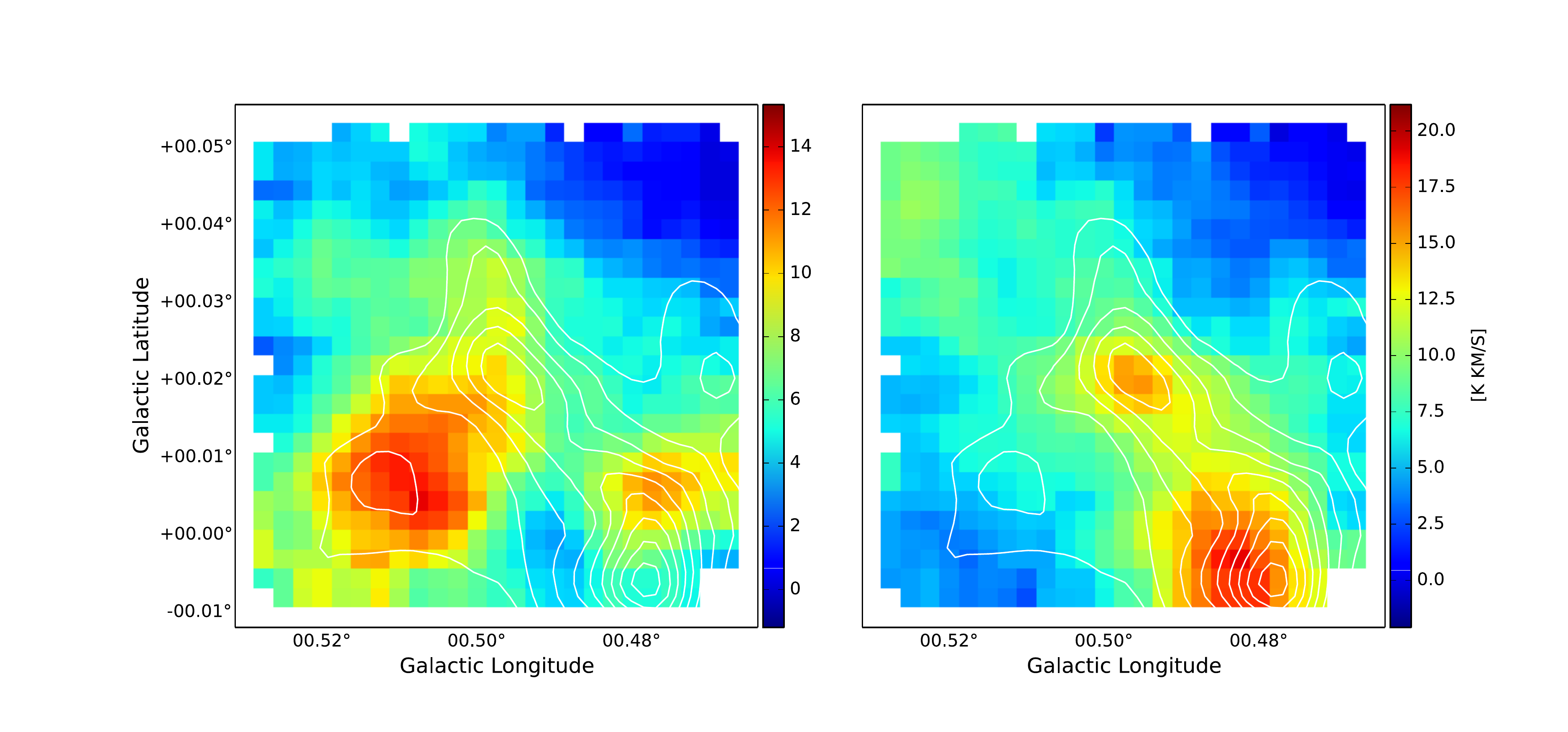}}
\hfill
\caption{{\bf{(a)}} Integrated HNCO intensity maps of cloud \cd \ showing emission integrated between --10 to 5 \kms \ ({\it{left}}) and 5 to 50 \kms \ ({\it{right}}). {\bf{(b)}} Integrated HNCO intensity maps over clouds \ce \ and \cf \ showing emission integrated between 5 -- 25 \kms \ ({\it{left}}) and 25 -- 50 \kms \ ({\it{right}}). Overlaid in white are contours from the HiGAL column density maps. These maps show that the emission over the lower velocity range does not match well the morphology of the dust emission toward these clouds. Thus, we assume that this emission arises from unrelated clouds along the line of sight. Note that HNCO emission is used here as it matches the HN$^{13}$C emission very closely but is detected with a much higher signal to noise.}
\end{figure}


If we are to investigate the cluster-forming potential of these clouds, we must first assess whether they are gravitationally bound. To do this we estimate and compare their dust and virial masses. Taking the radii and line-width estimates and assuming a uniform spherical density distribution, we use M$_{vir}$ = $k$R$\Delta$V$^{2}$ \citep[where $k$ = 126 and the units of M, R and $\Delta$V are M$_{\odot}$, pc and km s$^{-1}$, respectively; see][]{virial} to derive virial masses of 1.1 x 10$^{5}$ M$_{\odot}$, 7.3 x 10$^{4}$ M$_{\odot}$ and 6.9 x 10$^{4}$ M$_{\odot}$ for clouds \deaf, respectively. 

Under these assumptions, we compare the dust masses with the virial masses to obtain virial ratios of $\alpha = M_{\rm vir}/M_{\rm dust}=\{1.4,0.7,0.9\}$. It is clear that to within $\sim50\%$, $M_{\rm vir}\sim M_{\rm dust}$ in all cases. We therefore conclude that all three clouds are likely gravitationally bound. We note that the effect of magnetic pressure has not been accounted for in this analysis.

\subsection{Time-scales}\

Using the derived masses, radii and line-widths, we derive the sound crossing times, cloud crossing times and free-fall times for these clouds (see Table 1). Firstly we estimate the sound speed as $c_{s} = \sqrt{kT_{g}/\mu}$, where $k$ is the Boltzmann constant, $\mu$ is the mean molecular weight (taken to be 2.8$m_{H}$) and $T_{g}$ is the gas temperature (assumed to be 70 K in all cases, \citealt{Ao_gas_temp}, though we reiterate that this has only been measured for G0.253$+$0.016 -- we are assuming that this is also true for the other clouds. Any associated uncertainty in the sound speed is smaller as $c_{s}$ \ap $T^{0.5}$). This is then used to estimate the Mach number as $\mathcal{M} = \sigma/c_{s}$ (where $\sigma = \Delta V/2.355$). We estimate the sound crossing time as $t_{sc} = R/c_{s}$ and the cloud crossing time as $t_{cc} = R/\sigma$. Finally, we estimate the free-fall time as $t_{\rm ff}=\sqrt{3\pi/32G\rho}$, where $\rho$ has been estimated under the assumption of a uniform spherical distribution.

\section{Discussion}\

Table 1 displays the derived global properties for clouds G0.253$+$0.016, \cd, \ce \ and \cf, along with characteristic densities and time-scales for each cloud. These results highlight that all of these clouds are extremely massive and compact objects, containing \ap 10$^{5}$ M$_{\odot}$ within only a few parsecs, resulting in correspondingly high column and volume densities and short dynamical time-scales. Coupling these results with the fact that the clouds are close to virial equilibrium, the conclusion of \citet{Bricklets} is confirmed; that these clouds are all excellent YMC progenitor candidates. Intriguingly, there are no clouds yet known to exist in the rest of the Galaxy that have such extreme masses and densities that are not forming stars prodigiously \citep{Ginsburg_clouds, atlasgal_hii}. The fact that clouds as massive and dense as these currently exhibit minimal signs of active star formation may therefore have important implications. For example, it may suggest that star and cluster formation proceeds differently at the Galactic centre than in the disk \citep[see][and references therein]{snl_ymc}, or that we have simply caught the earliest stages of YMC formation in four separate cases. Given that the environment at the Galactic centre is extreme compared to the disk \citep{shetty_cmz,Diederik_highz}, it is plausible that star formation may be inhibited in some way \citep{snl_sf,Diederik_cmz_sf}. However, it is known that YMCs can and do form near the Galactic centre since at least two YMCs, the Arches and Quintuplet clusters, have formed there. The existence of YMCs at the Galactic centre therefore adds further weight to the conclusion that these extreme clouds have the capacity to form YMCs. We now investigate how these clouds can be used to gain insight in to the process of YMC formation.

\subsection{Comparing Clouds \& Clusters}
Having derived the global properties of G0.253$+$0.016, \deaf \ and confirming that they are sufficiently massive and dense to potentially form YMCs, we now compare their properties to the observed properties of the intermediate and final stages of YMC evolution. The reasoning here is that through comparing YMCs in their initial, intermediate, and final stages, we can begin to build up a coherent picture of how these different stages connect to one-another and ultimately determine whether the process of YMC formation is fundamentally different from that which forms low-mass clusters. We posit two cluster formation scenarios, which can be distinguished by comparing the initial gas sizes and densities ($R^{gas}_{init}$, $\rho^{gas}_{init}$) with those of the resultant stellar population ($R^{*}_{fin}$, $\rho^{*}_{fin}$) \citep[see][and references therein for a more detailed discussion]{snl_ymc}: 
\begin{indentpar}{1cm}
(i) A bound, centrally-condensed stellar population forms in an extremely compact natal gas cloud (i.e. $R^{gas}_{init}$ \textless \ $R^{*}_{fin}$; $\rho^{gas}_{init}$ \textgreater \ $\rho^{*}_{fin}$). Feedback processes gradually remove the remaining gas, diluting the global potential and causing the cluster to expand towards its final, un-embedded phase \citep[see e.g.][]{Lada_stellar_associations,cluster_mass_loss,nate_gas_removal,B&K_pop}. This results in a bound, spherical cluster with $R^{*}_{fin}$ \textgreater \ $R^{gas}_{init}$.

\

(ii) Stars and sub-clusters form in a gas cloud with $R^{gas}_{init}$ \textgreater \ $R^{*}_{fin}$, $\rho^{gas}_{init}$ \textless \ $\rho^{*}_{fin}$. They form throughout the spatial extent of their natal gas clouds, following the hierarchical structure of the interstellar medium \citep{ISM_structure}. A heightened star formation efficiency (SFE) in the densest peaks leads to gas exhaustion on local scales, causing stellar dynamics to eventually dominate \citep{Diederik_cluster_dynamics,cluster_fragmentation}. The subsequent hierarchical merging of these condensations results in a centrally concentrated, bound cluster \citep{ymc_merger_formation,Tiger14}.
\end{indentpar}

In essence, we should be able to distinguish between these two scenarios by studying the most likely progenitor systems -- massive and compact molecular clouds. If scenario (i) is a common mode of YMC formation, we ought to see $\gtrsim$ 10$^{5}$ M$_{\odot}$ clouds that are {\it{more or equally as compact}} as Galactic YMCs. If instead scenario (ii) is favoured, then we might expect to see such clouds with sizes larger than those of YMCs that show fragmented sub-structure on small spatial scales. Given that G0.253$+$0.016, \deaf \ represent the most extreme quiescent molecular clouds known in the Galaxy, they offer an ideal sample of progenitor systems with which we can investigate the validity of these two scenarios prior to the loss of initial structure due to feedback from high-mass star formation. We can begin to distinguish between these scenarios simply by looking at their mass surface density profiles to see how they compare to Galactic YMCs and proto-YMCs.

We use the HiGAL column density maps to obtain mass surface density profiles for clouds \deaf \ by calculating the enclosed mass (see \S {\it{2.2}}) within increasing circular apertures (centred on column density peaks) and dividing by the area of the corresponding aperture. For G0.253$+$0.016, we use ALMA Cycle 0 + single-dish data \citep{Jill_pdf_2014}. These data consist of ALMA observations at 3 mm, with an angular resolution of 1.7" (0.07 pc at a distance of 8.4 kpc). The data were then combined with Herschel 500 $\mu m$ to recover the flux that is filtered out by the interferometric observation. The resolution of the ALMA data allows us to investigate the surface density profile of this cloud down to much smaller radii. Interestingly, \citet{Brick_jill} do find evidence for fragmentation and hierarchical small-scale structure in G0.253$+$0.016. Furthermore, the ALMA data \citep[][]{Jill_pdf_2014} directly reveals such sub-structure within this cloud, showing that the gas is highly fragmented and contains a number of dense cores distributed throughout the cloud (we direct the reader to \citealt{rathborne15} for detailed analysis of the cloud's sub-structure). This suggests that the internal structure of G0.253$+$0.016 is consistent with that predicted by scenario (ii). Higher resolution data are required for the other clouds to assess whether this is true in all cases.

In all following plots, we have applied a multiplicative factor of 1/3 to the data for G0.253$+$0.016, \q{c}, \deaf \ such that we can infer what the resultant stellar population would look like if it were to form with a star formation efficiency (SFE, $\epsilon$) of 1/3 at the current mass distribution of the clouds. In reality, SFE will of course vary throughout the cloud, where it will be enhanced towards local density peaks -- this factor is chosen to represent a global SFE.

\subsubsection{Identifying the Intermediate Phase}

To investigate the active star-forming phase of YMC formation, we choose to study the gas {\it{and}} stellar content of the well-known star-forming Sagittarius B2 complex \citep[Sgr B2, e.g.][]{sgrb2}. We include both Sgr B2 Main and North, as these are potential proto-YMCs in a deeply embedded phase and lie in the same region as G0.253$+$0.016, \deaf. \citet{sgrb2} use data from the Submillimeter Array (SMA) to investigate the dense sub-structure within Sgr B2 Main and North. Interestingly, they find that Sgr B2 Main contains many sub-mm sources that appear to have a fragmented spatial distribution, whereas Sgr B2 North only contains two sub-mm sources. They propose that this may suggest that the main cluster is more evolved and that northern cluster is less evolved and characterised by monolithic high-mass star formation. 

The emission from the gas and dust in Sgr B2 is saturated in the HiGAL data. To obtain surface density profiles for the gas in Sgr B2 Main \& North, we instead use data from the Bolocam Galactic Plane Survey \citep[BGPS,][]{bgps1,bgps2,bgps3}. These data are at a wavelength of 1.1 mm and provide a pixel scale of 7.2'' (\ap 0.3 pc at a distance of 8.4 kpc).

As we are using these data to generate mass surface density profiles for the clouds, we convert the data from units of intensity to units of mass. We do this using the following relation (taken from \citealt{Jens_mass_in}, equation A.31, appendix A) --
\begin{equation}
\begin{split}
M = 0.12 M_{\odot}  \left( {\rm e}^{1.439 (\lambda / {\rm mm})^{-1}
      (T / {\rm 10 ~ K})^{-1}} - 1 \right) \\
   \cdot \left( \frac{\kappa_{\nu}}{0.01 \rm ~ cm^2 ~ g^{-1}} \right)^{-1} 
   \left( \frac{F_{\nu}}{\rm Jy} \right)
  \left( \frac{d}{\rm 100 ~ pc} \right)^2
  \left( \frac{\lambda}{\rm mm} \right)^{3},
\end{split}
\end{equation}

\noindent where $M$ is mass, $\lambda$ is wavelength, $T$ is the dust temperature, $k_{\nu}$ is the dust opacity, $F_{\nu}$ is the integrated flux and $d$ is distance. The dust temperature is assumed to be 20 K, though we note that a star-forming complex like Sgr B2 will not be isothermal due to heating from the embedded HII regions and stellar population. Additionally, it is known that the kinematic structure of Sgr B2 is complex, with multiple velocity components towards the region. As a result, any mass and density estimates for the gas in this region will be upper limits. Schmiedeke et al. (in prep) model the region in more detail. 

The only observationally unconstrained parameter in the above relation is the dust opacity ($k_{\nu}$). To estimate this, we use the following relation, given in \S 3.2 of \citet{Cara_higal} -- 
\begin{equation}
k_{\nu} = 0.04 \rm ~ cm^{2} ~ g^{-1} \left(\frac{\nu}{505 ~ GHz}\right)^{1.75},
\end{equation}

\noindent where $\nu$ is the frequency. Note that this contains the explicit assumption that the gas-to-dust ratio is 100.

Obtaining a mass surface density profile of the proto-cluster(s) within Sgr B2 is difficult due to the high column densities and hence extinction towards the region. To overcome this, we first take the positions and zero age main-sequence (ZAMS) spectral classifications of the stellar sources embedded in the Ultra-Compact HII (UCHII) regions from Tables 2 and 3 in \citet{Gaume}. We then convert the spectral type of each source to a representative mass using the spectroscopic masses of ZAMS OB stars given in Table 5 (column 8) of \citet{Vacca}. Knowing the spatial distribution and masses of the OB stars within Sgr B2 and assuming a distance of 8.4 kpc, we then proceed as previously and calculate the {\it{total}} stellar mass enclosed within increasing circular apertures. Given that the observations in \citet{Gaume} are sensitive only to stars $\gtrsim$ 10 M$_{\odot}$, we correct for the total mass by applying Kroupa-type IMF \citep{Kroupa_IMF}, for which the fraction of mass \textgreater 10 M$_{\odot}$ is \ap 0.16. The total mass is therefore estimated by applying a multiplicative factor of 1/0.16 and normalising by the mean stellar mass of \ap 0.5 M$_{\odot}$. Taking all 25 sources in \citet{Gaume}, we estimate a total stellar mass of \ap 3.5 x 10$^{3}$ M$_{\odot}$. The UCHII regions in Sgr B2 also indicate a sub-structured distribution of the embedded high mass stars -- possibly consistent with scenario (ii).  

\subsubsection{Comparison to Galactic YMCs}

For comparison with Galactic YMCs, we choose to discuss the Arches cluster as it is massive ($M$ \ap 2 x 10$^{4}$ M$_{\odot}$), compact ($R_{eff}$ \ap 0.4 pc), young (Age \ap 2 Myr) and is situated towards the Galactic centre and therefore a similar environment to Sgr B2, G0.253$+$0.016, \deaf \ \citep{ymc_port}. If YMC radii and central densities are related to the tidal radius, such that more compact YMCs are formed in stronger tidal fields, it is important that we compare clouds and clusters within the Galactic centre so as to eliminate any environmental variations. The Quintuplet cluster is also situated towards the Galactic centre. However, it is an older system than the Arches and it has been suggested that the disruption time-scale of clusters at the Galactic centre is short, occurring over $\sim$10 Myr \citep{kim_gc_disrup, pz_gc_disruption,Diederik_cmz_sf}. As such, we use only the Arches for comparison here as it is more likely representative of an initial YMC distribution. We utilise the observed surface density profile shown in Figure 16 of \citet{Arches_esp} along with the given best-fit King density profile and parameters, correcting for the number and masses of stars below 10 M$_{\odot}$ with a Kroupa-type IMF. We also use the observed cumulative mass profile given in Figure 3 of \citet{Arches_pz}, again IMF-corrected for stars below 10 M$_{\odot}$.


\

Figure 6 displays the enclosed mass as a function of radius for G0.253$+$0.016, \q{d}, \q{e} and \q{f} (solid lines) given $\epsilon$ = 1/3, the gas in Sagittarius B2 Main and North (dashed lines) and the Arches cluster from \citet{Arches_esp} (black, dash/dot) and \citet{Arches_pz} (red, dash/dot). The plot shows that G0.253$+$0.016, \deaf \ all contain enough mass such that they can form a YMC of $M$ \textgreater 10$^{4}$ M$_{\odot}$ given a SFE of 1/3. However, it is clear in {\it{all}} cases that the distribution of mass is much less centrally concentrated than in the Arches, leading us to conclude that these clouds cannot form an Arches-like YMC at their current densities. Unless they were to somehow condense rapidly on global scales within a free-fall time (\ap 0.5 Myr) before the onset of any widespread star formation, it seems implausible that these clouds will form clusters in accordance with scenario (i). We see that this is also true of the gas content of Sgr B2 Main and North. Despite the high mass and density of these regions, the gas is too extended on global scales.

Figure 7 shows the resultant mass surface density profiles as a function of radius for G0.253$+$0.016, \q{d}, \q{e} and \q{f} given $\epsilon$ = 1/3 (solid lines), the gas in Sagittarius B2 Main and North (dashed lines), the proto-cluster(s) embedded within Sagittarius B2 Main and North as calculated from the UCHII region distribution (blue and red dash/dot, respectively) and the Arches cluster overlaid with the fit from \citet{Arches_esp} (black open circles, dash/dot). We again see that G0.253$+$0.016, \deaf \ are less centrally concentrated than the Arches' stellar distribution -- further dynamical evolution would be required to condense any resultant stellar populations if they were to form at the current densities of the clouds. We see that this is also true for the gas content of Sgr B2 Main and North. Even though they are significantly more massive, dense and evolved than the other clouds, the gas is still smoothly distributed on global scales. This is in contrast to the inferred stellar content in Sgr B2, where the HII regions of both Sgr B2 Main and North are concentrated in a small volume (\textless \ 0.1 pc). This shows that dense, centrally concentrated proto-clusters are able to form in clouds that are not very centrally concentrated, as predicted by hydrodynamical cluster formation simulations \citep{Diederik_cluster_dynamics}. Furthermore, we find that, at least for this sample of clouds towards the Galactic centre, the final stellar distribution of an Arches-like YMC is more compact than the global distribution of stars {\it{and}} gas at any prior stage of the formation process. Given that these are the most massive and dense quiescent clouds yet found in the Galaxy, we therefore conclude that scenario (i) is disfavoured as a likely mode of YMC formation. 

Having investigated the absolute mass and surface density profiles, we now wish to identify the {\it{shape}} of the radial mass distribution so as to investigate any differences. Figure 8 displays the normalised mass surface density profiles for G0.253$+$0.016, \q{d}, \q{e} and \q{f} (solid lines), the gas in Sagittarius B2 Main and North (dashed lines), the Sagittarius B2 Main and North proto-clusters (blue and red dash/dot, respectively) and the Arches cluster (black, dash/dot), where all profiles have been normalised to unity at a radius of \ap 0.3 pc (the resolution of the BGPS data at a distance of 8.4 kpc). This plot shows that in {\it{all}} cases, the gas in the YMC progenitors is over-dense at large radii and under-dense at small radii, compared to the stellar components in both the Arches cluster and the Sgr B2 proto-clusters. This suggests that if these clouds are to form YMCs, the resultant stellar population would have to dynamically interact such that it would relax into a much more centrally-condensed distribution. It is interesting that the gas distribution in both Sagittarius B2 Main and North on global scales is very similar to that in the quiescent \q{dust-ridge} clouds. We see evidence for clustered massive star formation in Sgr B2, yet the global gas content looks identically distributed to that in a quiescent cloud. This is in conflict with scenario (i), which requires the gas to be more centrally concentrated prior to the formation of a YMC.

It is also interesting to note the high mass surface density of Sgr B2 proto-clusters on scales smaller than \ap 0.1 pc. The bulk of the stellar mass, though roughly an order of magnitude lower than the Arches, is highly concentrated within a small core region. It is difficult to determine whether this is consistent with scenario (i), in which case the stars may have formed in a centrally-condensed distribution, or whether it is consistent with scenario (ii), where the stars may be distributed in this way as a result of rapid dynamical interaction.

\begin{table*}
\begin{center}
  \label{tab:global_properties}
  \begin{tabular}{ccccccccccccccc}
    \hline
    Cloud	& M          & D$^{\dagger}$     & R     & T$_{\rm dust}$ & $\Delta$V & M$_{vir}$ & $\alpha$ & $n$     & N$_{H_2}$  & $t_{cc}$ & $t_{sc}$ & $t_{ff}$ & $\mathcal{M}$ & R$_{5}$\\
    & 	10$^{4}$ M$_\odot$ & kpc & pc & K          & km/s    & 10$^{4}$ M$_\odot$ & -- &10$^{4}$ cm$^{-3}$ & 10$^{24}$ cm$^{-2}$ & Myr   & Myr   & Myr & -- & pc \\ \hline
G0.25 &11.9   & 8.4   & 2.9  & 19-27        & 15.1 $\pm$ 1.0$^{*}$     & 8.3  & 0.7 & 1.7 & 0.6  &  0.44   & 6.2  &  0.51 & 14.1 & 1.0\\ 
    d  	&7.6     & 8.4   & 3.2   & 19-23        & 16.3 $\pm$ 1.5      & 10.7 & 1.4 & 0.8      & 0.3    &  0.45   & 6.9        & 0.74 & 15.2 & 1.2\\ 
    e 	&11.2      & 8.4   & 2.4   & 17-22        & 15.5 $\pm$ 0.5   & 7.3   & 0.7 & 2.8    & 0.9    &  0.36   & 5.2        & 0.40 & 14.5 & 0.8\\ 
    f   &7.3       & 8.4   & 2.0   & 18-22        & 16.5 $\pm$ 3.2      & 6.9 & 0.9 & 3.2      & 0.8     &  0.28   & 4.3        & 0.37 & 15.4 & 1.0\\ \hline \hline
  \end{tabular}
  \caption{Global properties of clouds G0.253$+$0.016, 'd', 'e' and 'f'. The columns show mass (M),
    distance (D), radius (R), dust temperature (T$_{\rm dust}$),
    linewidth ($\Delta$V), Virial mass (M$_{vir}$), Virial ratio ($\alpha$) average volume density ($n$), average column density
    (N$_{H_2}$), cloud-crossing time ($t_{cc}$), sound-crossing time
    ($t_{sc}$), free-fall time ($t_{ff}$), Mach number ($\mathcal{M}$) and the radius within which 5 x 10$^{4}$ M$_{\odot}$ is enclosed (R$_{5}$). *{\it{Result from \citet{Brick}. $^{\dagger}$Galactrocentric distance estimate from \citet{distance}  -- all clouds are assumed to be at this distance. Sound speed is calculated using a gas temperature of 70 K \citep{Ao_gas_temp}.}}}
\end{center}
\end{table*}

\begin{figure*}
\begin{center}
\includegraphics[scale=0.9,angle=-90]{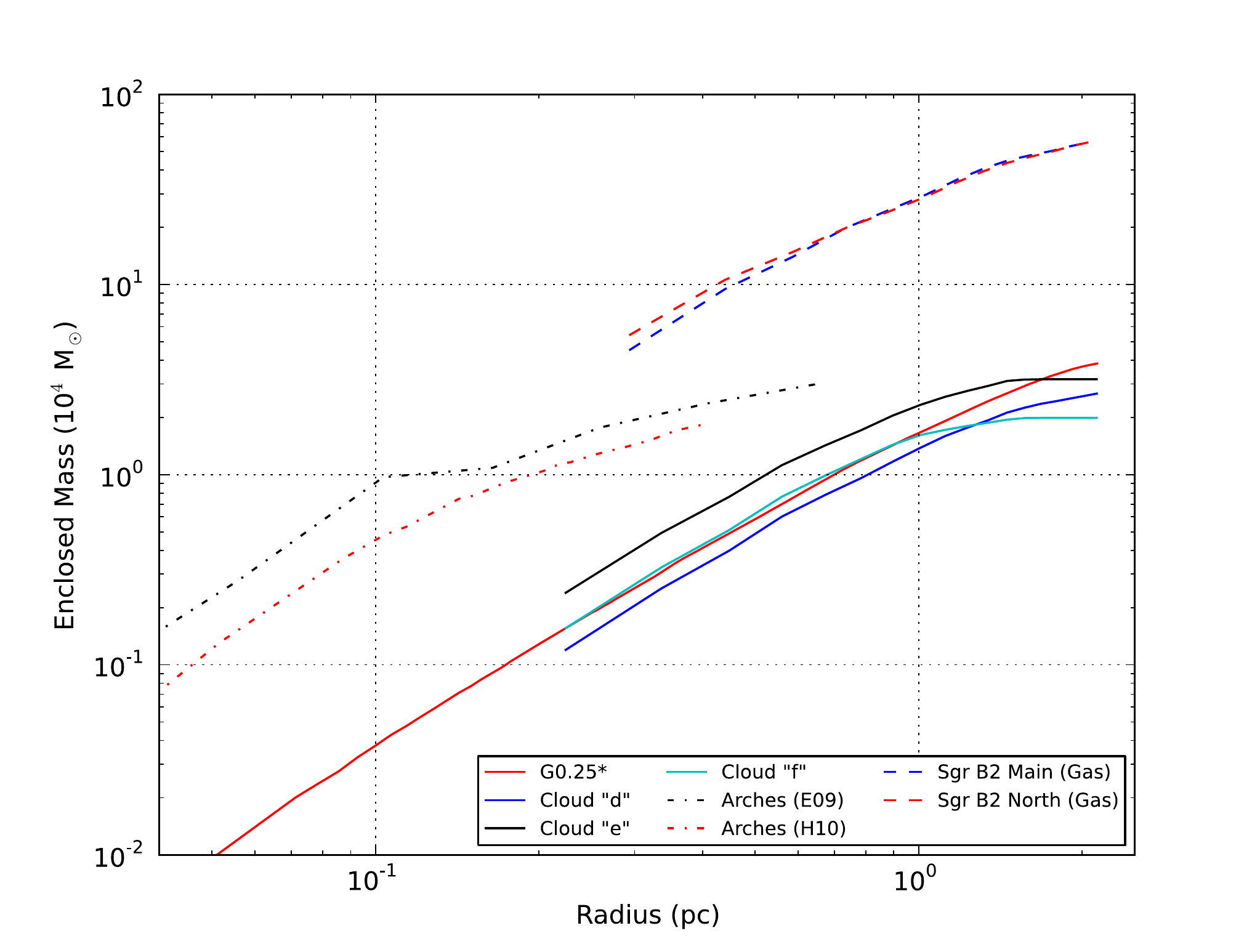} \\
 \caption{Enclosed mass as a function of radius for G0.253$+$0.016, \q{d}, \q{e}, \q{f} given $\epsilon$ = 1/3 (solid lines), the gas in Sagittarius B2 Main and North (dashed lines) and the Arches cluster from \citet{Arches_esp} (E09; black, dash/dot) and \citet{Arches_pz} (H10; red, dash/dot). G0.253$+$0.016* (red) ALMA cycle 0 + single-dish data \citep{Jill_pdf_2014}. We see that in all cases, the gas in the proto-cluster clouds has a more extended distribution of mass, whereas the stellar population of the Arches is more centrally concentrated.}
\end{center}
\end{figure*}

\begin{figure*}
\begin{center}
\includegraphics[scale=0.9,angle=-90]{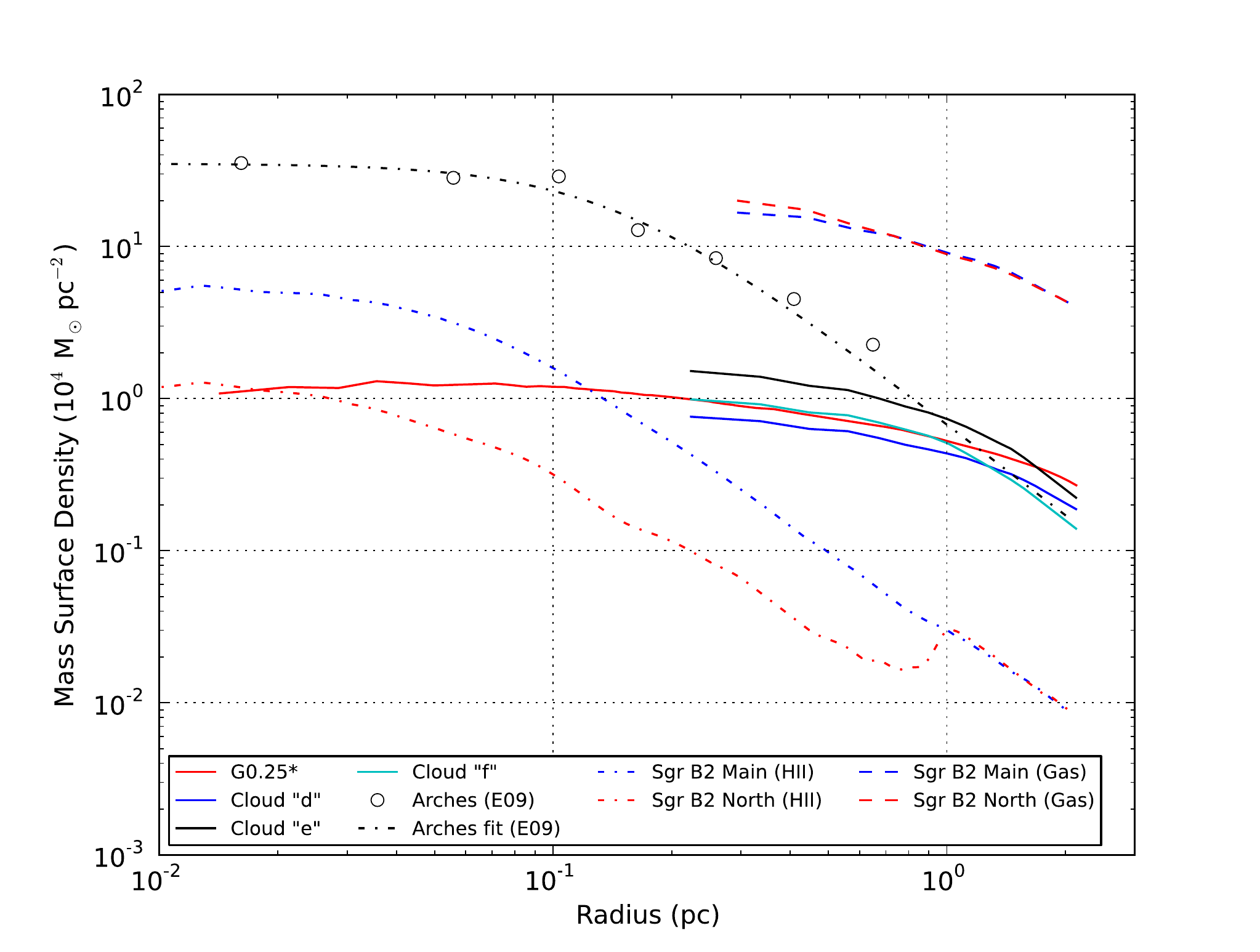} \\
 \caption{Mass surface density profiles as a function of radius for G0.253$+$0.016, \q{d}, \q{e}, \q{f} given $\epsilon$ = 1/3 (solid lines), the gas in Sagittarius B2 Main and North (dashed lines), the proto-cluster(s) embedded within Sagittarius B2 Main and North as calculated from the UCHII region distribution (blue and red dash/dot, respectively) and Arches cluster overlaid with fit from \citet{Arches_esp} (E09; black open circles, dash/dot). These profiles show that the final stellar distribution of the YMC is {\it{more compact}} than the global distribution of stars and gas at any prior stage of the formation process. Note that the variation of the profiles for the HII region distribution in Sgr B2 at small radii is a result of the small number of sources detected there. The bump at R \ap 1.0 pc is due to Sgr B2 Main entering the aperture at that radius. G0.253$+$0.016* (red) ALMA cycle 0 + single-dish data \citep{Jill_pdf_2014}.}
\end{center}
\end{figure*}

\begin{figure*}
\begin{center}
\includegraphics[scale=0.9,angle=-90]{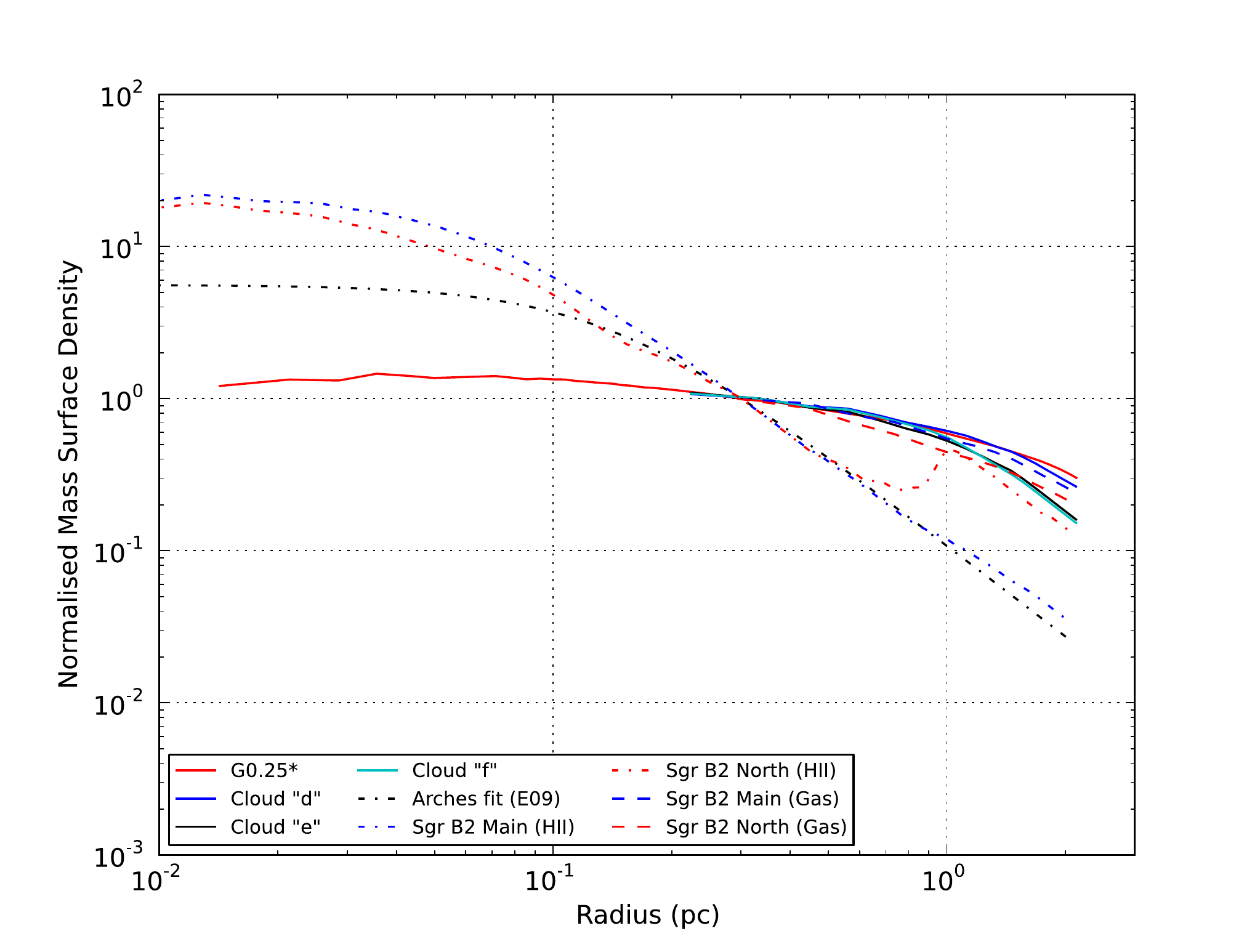} \\
 \caption{Normalised mass surface density profiles as a function of radius for the gas in G0.253$+$0.016, \q{d}, \q{e}, \q{f} (solid lines), the gas in Sagittarius B2 Main and North (dashed lines), the proto-cluster(s) embedded within Sagittarius B2 Main and North as calculated from the UCHII region distribution (blue and red dash/dot, respectively) and the Arches cluster from \citet{Arches_esp} (E09; black dash/dot). Each profile has been normalised to unity at a radius of \ap 0.3 pc, as this is the resolution of the BGPS data. G0.253$+$0.016* (red) ALMA cycle 0 + single-dish data \citep{Jill_pdf_2014}. The shapes of the profiles at larger radii suggests an evolutionary progression on global scales from a dispersed natal gas cloud to a centrally condensed stellar cluster as a function of star formation activity.}
\end{center}
\end{figure*}

\section{Conclusions}\

Using HiGAL far-IR continuum data and MALT90 millimetre spectral line data, we derive the global properties of four likely YMC precursors towards the Galactic centre. We find that these clouds, G0.253$+$0.016, \deaf, are all high mass ($M$ \ap 10$^{5}$ M$_{\odot}$), cold ($T_{dust}$ \ap 20 K) and dense ($\rho$ \ap 10$^{4}$ cm$^{-3}$). They are all close to virial equilibrium and are therefore likely to be gravitationally bound. These results confirm that they are excellent YMC progenitor candidates -- they are amongst the most massive and dense molecular clouds known to exist in the Galaxy, yet they are devoid of any widespread star formation.

Comparison of the mass surface density distributions of these clouds with the Sagittarius B2 proto-YMC and the Arches YMC shows that these clouds are not compact or centrally concentrated enough to form an Arches-like YMC in their current state. If they are to form YMCs, dynamical evolution during the early formation must further condense the resultant stellar population. Furthermore, we find that the stellar content of both Sagittarius B2 Main and North is significantly more centrally concentrated than the global gas content. This shows that dense, centrally concentrated stellar clusters can form from gas that is not very centrally condensed, thus disfavouring a \q{monolithic collapse} mode of YMC formation in which gas expulsion causes the YMC to end up less compact than the progenitor gas cloud.

\section*{Acknowledgements}

The authors would like to thank the anonymous referee for their constructive comments. Daniel Walker would like to thank both Richard Parker and Adam Ginsburg for their insightful comments. This paper makes use of the following ALMA data:  ADS/JAO.ALMA\#2011.0.00217.S. ALMA is a partnership of the European Southern Observatory (ESO) representing member states, Associated Universities Incorporated (AUI) and the National radio Astronomy Observatories (NRAO) for the National Science Foundation (NSF) in the USA, NINS  in Japan, NRC in Canada, and NSC and ASIAA in Taiwan, in cooperation with the Republic of Chile.  The Joint ALMA Observatory (JAO) is operated by ESO (Europe) , AUI/NRAO  (USA), and NAOJ (Japan). The National Radio Astronomy Observatory is a facility of the National Science Foundation operated under cooperative agreement by Associated Universities, Inc. This research made use of the NASA Astrophysical Data System.

\bibliography{dw15_v3.0}

\begin{thebibliography}{66}
\expandafter\ifx\csname natexlab\endcsname\relax\def\natexlab#1{#1}\fi

\bibitem[{{Aguirre} {et~al}\mbox{.}(2011){Aguirre}, {Ginsburg}, {Dunham}, \&
  {et al}}]{bgps2}
{Aguirre} J.~E., {Ginsburg} A.~G., {Dunham} M.~K., {et al}, 2011, \apjs, 192, 4

\bibitem[{{Ao} {et~al}\mbox{.}(2013){Ao}, {Henkel}, {Menten}, \& {et
  al}}]{Ao_gas_temp}
{Ao} Y., {Henkel} C., {Menten} K.~M., {et al}, 2013, \aap, 550, A135

\bibitem[{{Bastian} \& {Goodwin}(2006)}]{nate_gas_removal}
{Bastian} N., {Goodwin} S.~P., 2006, \mnras, 369, L9

\bibitem[{{Battersby} {et~al}\mbox{.}(2011){Battersby}, {Bally}, {Ginsburg}, \&
  {et al}}]{Cara_higal}
{Battersby} C., {Bally} J., {Ginsburg} A., {et al}, 2011, \aap, 535, A128

\bibitem[{{Baumgardt} \& {Kroupa}(2007)}]{B&K_pop}
{Baumgardt} H., {Kroupa} P., 2007, \mnras, 380, 1589

\bibitem[{{Boily} \& {Kroupa}(2003)}]{cluster_mass_loss}
{Boily} C.~M., {Kroupa} P., 2003, \mnras, 338, 673

\bibitem[{{Caswell} {et~al}\mbox{.}(2010){Caswell}, {Fuller}, {Green}, \& {et
  al}}]{c_maser}
{Caswell} J.~L., {Fuller} G.~A., {Green} J.~A., {et al}, 2010, \mnras, 404,
  1029

\bibitem[{{Clark} {et~al}\mbox{.}(2005){Clark}, {Negueruela}, {Crowther}, \&
  {Goodwin}}]{Clark_west}
{Clark} J.~S., {Negueruela} I., {Crowther} P.~A., {Goodwin} S.~P., 2005, \aap,
  434, 949

\bibitem[{{Davies} {et~al}\mbox{.}(2007){Davies}, {Figer}, {Kudritzki}, \& {et
  al}}]{Ben_RSG}
{Davies} B., {Figer} D.~F., {Kudritzki} R.-P., {et al}, 2007, \apj, 671, 781

\bibitem[{{Elmegreen} \& {Efremov}(1997)}]{elm_gcs}
{Elmegreen} B.~G., {Efremov} Y.~N., 1997, \apj, 480, 235

\bibitem[{{Espinoza}, {Selman} \& {Melnick}(2009){Espinoza}, {Selman}, \&
  {Melnick}}]{Arches_esp}
{Espinoza} P., {Selman} F.~J., {Melnick} J., 2009, \aap, 501, 563

\bibitem[{{Figer} {et~al}\mbox{.}(1999){Figer}, {Kim}, {Morris}, {Serabyn},
  {Rich}, \& {McLean}}]{Figer_arches}
{Figer} D.~F., {Kim} S.~S., {Morris} M., {Serabyn} E., {Rich} R.~M., {McLean}
  I.~S., 1999, \apj, 525, 750

\bibitem[{{Forster} \& {Caswell}(1999)}]{c_watermaser}
{Forster} J.~R., {Caswell} J.~L., 1999, \aaps, 137, 43

\bibitem[{{Foster} {et~al}\mbox{.}(2011){Foster}, {Jackson}, {Barnes}, \& {et
  al}}]{malt90_foster_11}
{Foster} J.~B., {Jackson} J.~M., {Barnes} P.~J., {et al}, 2011, \apjs, 197, 25

\bibitem[{{Foster} {et~al}\mbox{.}(2013){Foster}, {Rathborne}, {Sanhueza}, \&
  {et al}}]{malt90_foster_13}
{Foster} J.~B., {Rathborne} J.~M., {Sanhueza} P., {et al}, 2013, \pasa, 30, 38

\bibitem[{{Fujii}, {Saitoh} \& {Portegies Zwart}(2012){Fujii}, {Saitoh}, \&
  {Portegies Zwart}}]{ymc_merger_formation}
{Fujii} M.~S., {Saitoh} T.~R., {Portegies Zwart} S.~F., 2012, \apj, 753, 85

\bibitem[{{Gaume} {et~al}\mbox{.}(1995){Gaume}, {Claussen}, {de Pree}, \& {et
  al}}]{Gaume}
{Gaume} R.~A., {Claussen} M.~J., {de Pree} C.~G., {et al}, 1995, \apj, 449, 663

\bibitem[{{Ginsburg} {et~al}\mbox{.}(2012){Ginsburg}, {Bressert}, {Bally}, \&
  {Battersby}}]{Ginsburg_clouds}
{Ginsburg} A., {Bressert} E., {Bally} J., {Battersby} C., 2012, \apjl, 758, L29

\bibitem[{{Ginsburg} {et~al}\mbox{.}(2013){Ginsburg}, {Glenn}, {Rosolowsky}, \&
  {et al}}]{bgps3}
{Ginsburg} A., {Glenn} J., {Rosolowsky} E., {et al}, 2013, \apjs, 208, 14

\bibitem[{{Girichidis} {et~al}\mbox{.}(2012){Girichidis}, {Federrath},
  {Banerjee}, \& {Klessen}}]{cluster_fragmentation}
{Girichidis} P., {Federrath} C., {Banerjee} R., {Klessen} R.~S., 2012, \mnras,
  420, 613

\bibitem[{{Goddard}, {Bastian} \& {Kennicutt}(2010){Goddard}, {Bastian}, \&
  {Kennicutt}}]{embedded_survival}
{Goddard} Q.~E., {Bastian} N., {Kennicutt} R.~C., 2010, \mnras, 405, 857

\bibitem[{{Goss} \& {Radhakrishnan}(1969)}]{NGC3603}
{Goss} W.~M., {Radhakrishnan} V., 1969, \aplett, 4, 199

\bibitem[{{Harfst}, {Portegies Zwart} \& {Stolte}(2010){Harfst}, {Portegies
  Zwart}, \& {Stolte}}]{Arches_pz}
{Harfst} S., {Portegies Zwart} S., {Stolte} A., 2010, \mnras, 409, 628

\bibitem[{{Holtzman} {et~al}\mbox{.}(1992){Holtzman}, {Faber}, {Shaya},
  {Lauer}, \& {et al}}]{Holt}
{Holtzman} J.~A., {Faber} S.~M., {Shaya} E.~J., {Lauer} T.~R., {et al}, 1992,
  \aj, 103, 691

\bibitem[{{Immer} {et~al}\mbox{.}(2012){Immer}, {Menten}, {Schuller}, \&
  {Lis}}]{Immer}
{Immer} K., {Menten} K.~M., {Schuller} F., {Lis} D.~C., 2012, \aap, 548, A120

\bibitem[{{Jackson} {et~al}\mbox{.}(2013){Jackson}, {Rathborne}, {Foster}, \&
  {et al}}]{malt90}
{Jackson} J.~M., {Rathborne} J.~M., {Foster} J.~B., {et al}, 2013, \pasa, 30,
  57

\bibitem[{{Johnston} {et~al}\mbox{.}(2014){Johnston}, {Beuther}, {Linz},
  {Schmiedeke}, {Ragan}, \& {Henning}}]{Brick_KJ}
{Johnston} K.~G., {Beuther} H., {Linz} H., {Schmiedeke} A., {Ragan} S.~E.,
  {Henning} T., 2014, \aap, 568, A56

\bibitem[{{Kauffmann} {et~al}\mbox{.}(2008){Kauffmann}, {Bertoldi}, {Bourke},
  {Evans}, \& {Lee}}]{Jens_mass_in}
{Kauffmann} J., {Bertoldi} F., {Bourke} T.~L., {Evans}, II N.~J., {Lee} C.~W.,
  2008, \aap, 487, 993

\bibitem[{{Kauffmann}, {Pillai} \& {Zhang}(2013){Kauffmann}, {Pillai}, \&
  {Zhang}}]{Brick_jens}
{Kauffmann} J., {Pillai} T., {Zhang} Q., 2013, \apjl, 765, L35

\bibitem[{{Kim}, {Morris} \& {Lee}(1999){Kim}, {Morris}, \&
  {Lee}}]{kim_gc_disrup}
{Kim} S.~S., {Morris} M., {Lee} H.~M., 1999, \apj, 525, 228

\bibitem[{{Kroupa}(2001)}]{Kroupa_IMF}
{Kroupa} P., 2001, \mnras, 322, 231

\bibitem[{{Kruijssen}(2012)}]{Diederik_cfe}
{Kruijssen} J.~M.~D., 2012, \mnras, 426, 3008

\bibitem[{{Kruijssen}(2014)}]{Diederik_submitted}
{Kruijssen} J.~M.~D., 2014, Classical and Quantum Gravity, 31, 244006

\bibitem[{{Kruijssen}, {Dale} \& {Longmore}(2015){Kruijssen}, {Dale}, \&
  {Longmore}}]{Diederik_orbit}
{Kruijssen} J.~M.~D., {Dale} J.~E., {Longmore} S.~N., 2015, \mnras, 447, 1059

\bibitem[{{Kruijssen} \& {Longmore}(2013)}]{Diederik_highz}
{Kruijssen} J.~M.~D., {Longmore} S.~N., 2013, \mnras, 435, 2598

\bibitem[{{Kruijssen} {et~al}\mbox{.}(2014){Kruijssen}, {Longmore},
  {Elmegreen}, \& {et al}}]{Diederik_cmz_sf}
{Kruijssen} J.~M.~D., {Longmore} S.~N., {Elmegreen} B.~G., {et al}, 2014,
  \mnras, 440, 3370

\bibitem[{{Kruijssen} {et~al}\mbox{.}(2012){Kruijssen}, {Maschberger},
  {Moeckel}, \& {et al}}]{Diederik_cluster_dynamics}
{Kruijssen} J.~M.~D., {Maschberger} T., {Moeckel} N., {et al}, 2012, \mnras,
  419, 841

\bibitem[{{Lada} \& {Lada}(2003)}]{Lada}
{Lada} C.~J., {Lada} E.~A., 2003, \araa, 41, 57

\bibitem[{{Lada}, {Margulis} \& {Dearborn}(1984){Lada}, {Margulis}, \&
  {Dearborn}}]{Lada_stellar_associations}
{Lada} C.~J., {Margulis} M., {Dearborn} D., 1984, \apj, 285, 141

\bibitem[{{Larson}(1981)}]{ISM_structure}
{Larson} R.~B., 1981, \mnras, 194, 809

\bibitem[{{Lis} {et~al}\mbox{.}(1999){Lis}, {Li}, {Dowell}, \&
  {Menten}}]{Bricklets_lis}
{Lis} D.~C., {Li} Y., {Dowell} C.~D., {Menten} K.~M., 1999, in ESA Special
  Publication, Vol. 427, The Universe as Seen by ISO, {Cox} P., {Kessler} M.,
  eds., p. 627

\bibitem[{{Lis} \& {Menten}(1998)}]{Lis_brick2}
{Lis} D.~C., {Menten} K.~M., 1998, \apj, 507, 794

\bibitem[{{Lis} {et~al}\mbox{.}(1994){Lis}, {Menten}, {Serabyn}, \&
  {Zylka}}]{Lis_brick1}
{Lis} D.~C., {Menten} K.~M., {Serabyn} E., {Zylka} R., 1994, \apjl, 423, L39

\bibitem[{{Longmore} {et~al}\mbox{.}(2013{\natexlab{a}}){Longmore}, {Bally},
  {Testi}, \& {et al}}]{snl_sf}
{Longmore} S.~N., {Bally} J., {Testi} L., {et al}, 2013{\natexlab{a}}, \mnras,
  429, 987

\bibitem[{{Longmore} {et~al}\mbox{.}(2013{\natexlab{b}}){Longmore},
  {Kruijssen}, {Bally}, \& {et al}}]{Bricklets}
{Longmore} S.~N., {Kruijssen} J.~M.~D., {Bally} J., {et al},
  2013{\natexlab{b}}, \mnras, 433, L15

\bibitem[{{Longmore} {et~al}\mbox{.}(2014){Longmore}, {Kruijssen}, {Bastian},
  \& {et al}}]{snl_ymc}
{Longmore} S.~N., {Kruijssen} J.~M.~D., {Bastian} N., {et al}, 2014,
  Protostars~\&~Planets~VI, in press, arXiv:1401.4175

\bibitem[{{Longmore} {et~al}\mbox{.}(2012){Longmore}, {Rathborne}, {Bastian},
  \& {et al}}]{Brick}
{Longmore} S.~N., {Rathborne} J., {Bastian} N., {et al}, 2012, \apj, 746, 117

\bibitem[{{MacLaren}, {Richardson} \& {Wolfendale}(1988){MacLaren},
  {Richardson}, \& {Wolfendale}}]{virial}
{MacLaren} I., {Richardson} K.~M., {Wolfendale} A.~W., 1988, \apj, 333, 821

\bibitem[{{McMullin} {et~al}\mbox{.}(2007){McMullin}, {Waters}, {Schiebel},
  {Young}, \& {Golap}}]{casa}
{McMullin} J.~P., {Waters} B., {Schiebel} D., {Young} W., {Golap} K., 2007, in
  Astronomical Society of the Pacific Conference Series, Vol. 376, Astronomical
  Data Analysis Software and Systems XVI, {Shaw} R.~A., {Hill} F., {Bell}
  D.~J., eds., p. 127

\bibitem[{{Molinari} {et~al}\mbox{.}(2011){Molinari}, {Bally},
  {Noriega-Crespo}, \& {et al}}]{Molinari_ring}
{Molinari} S., {Bally} J., {Noriega-Crespo} A., {et al}, 2011, \apjl, 735, L33

\bibitem[{{Molinari} {et~al}\mbox{.}(2010){Molinari}, {Swinyard}, {Bally}, \&
  {et al}}]{higal}
{Molinari} S., {Swinyard} B., {Bally} J., {et al}, 2010, \pasp, 122, 314

\bibitem[{{Morris} \& {Serabyn}(1996)}]{CMZ}
{Morris} M., {Serabyn} E., 1996, \araa, 34, 645

\bibitem[{{Parker} {et~al}\mbox{.}(2014){Parker}, {Wright}, {Goodwin}, \&
  {Meyer}}]{Tiger14}
{Parker} R.~J., {Wright} N.~J., {Goodwin} S.~P., {Meyer} M.~R., 2014, \mnras,
  438, 620

\bibitem[{{Portegies Zwart} {et~al}\mbox{.}(2002){Portegies Zwart}, {Makino},
  {McMillan}, \& {Hut}}]{pz_gc_disruption}
{Portegies Zwart} S.~F., {Makino} J., {McMillan} S.~L.~W., {Hut} P., 2002,
  \apj, 565, 265

\bibitem[{{Portegies Zwart}, {McMillan} \& {Gieles}(2010){Portegies Zwart},
  {McMillan}, \& {Gieles}}]{ymc_port}
{Portegies Zwart} S.~F., {McMillan} S.~L.~W., {Gieles} M., 2010, \araa, 48, 431

\bibitem[{{Qin} {et~al}\mbox{.}(2011){Qin}, {Schilke}, {Rolffs}, \& {et
  al}}]{sgrb2}
{Qin} S.-L., {Schilke} P., {Rolffs} R., {et al}, 2011, \aap, 530, L9

\bibitem[{{Rathborne} {et~al}\mbox{.}(2014{\natexlab{a}}){Rathborne},
  {Longmore}, {Jackson}, \& {et al}}]{Brick_jill}
{Rathborne} J.~M., {Longmore} S.~N., {Jackson} J.~M., {et al},
  2014{\natexlab{a}}, \apj, 786, 140

\bibitem[{{Rathborne} {et~al}\mbox{.}(2014{\natexlab{b}}){Rathborne},
  {Longmore}, {Jackson}, \& {et al}}]{Jill_pdf_2014}
{Rathborne} J.~M., {Longmore} S.~N., {Jackson} J.~M., {et al},
  2014{\natexlab{b}}, \apjl, 795, L25

\bibitem[{{Rathborne} {et~al}\mbox{.}(2015){Rathborne}, {Longmore}, {Jackson},
  \& {et al}}]{rathborne15}
{Rathborne} J.~M., {Longmore} S.~N., {Jackson} J.~M., {et al}, 2015,
  \apj~submitted

\bibitem[{{Reid} {et~al}\mbox{.}(2009){Reid}, {Menten}, {Zheng}, \& {et
  al}}]{distance}
{Reid} M.~J., {Menten} K.~M., {Zheng} X.~W., {et al}, 2009, \apj, 700, 137

\bibitem[{{Rosolowsky} {et~al}\mbox{.}(2010){Rosolowsky}, {Dunham}, {Ginsburg},
  \& {et al}}]{bgps1}
{Rosolowsky} E., {Dunham} M.~K., {Ginsburg} A., {et al}, 2010, \apjs, 188, 123

\bibitem[{{Shetty} {et~al}\mbox{.}(2012){Shetty}, {Beaumont}, {Burton},
  {Kelly}, \& {Klessen}}]{shetty_cmz}
{Shetty} R., {Beaumont} C.~N., {Burton} M.~G., {Kelly} B.~C., {Klessen} R.~S.,
  2012, \mnras, 425, 720

\bibitem[{{Silva-Villa} \& {Larsen}(2011)}]{silva_cfe}
{Silva-Villa} E., {Larsen} S.~S., 2011, \aap, 529, A25

\bibitem[{{Urquhart} {et~al}\mbox{.}(2013){Urquhart}, {Moore}, {Schuller}, \&
  {et al}}]{atlasgal_hii}
{Urquhart} J.~S., {Moore} T.~J.~T., {Schuller} F., {et al}, 2013, \mnras, 431,
  1752

\bibitem[{{Vacca}, {Garmany} \& {Shull}(1996){Vacca}, {Garmany}, \&
  {Shull}}]{Vacca}
{Vacca} W.~D., {Garmany} C.~D., {Shull} J.~M., 1996, \apj, 460, 914

\bibitem[{{Whitmore}(2002)}]{Whitmore}
{Whitmore} B.~C., 2002, in IAU Symposium, Vol. 207, Extragalactic Star
  Clusters, {Geisler} D.~P., {Grebel} E.~K., {Minniti} D., eds., p. 367

\end{thebibliography}

\bsp

\label{lastpage}

\end{document}